\documentclass[11pt]{article}
\usepackage{amsfonts,amssymb,amsmath}

\newcommand{\ket}[1]{\mbox{$| #1 \rangle$}}

\newcommand{\braket}[2]{\mbox{$\langle #1 | #2 \rangle$}}

\def\squareforqed{\hbox{\rlap{$\sqcap$}$\sqcup$}}
\def\qed{\ifmmode\squareforqed\else{\unskip\nobreak\hfil
\penalty50\hskip1em\null\nobreak\hfil\squareforqed
\parfillskip=0pt\finalhyphendemerits=0\endgraf}\fi}

\DeclareMathSymbol{\leqslant}{\mathrel}{AMSa}{"36}  

\newcommand{\complex}{\mathbb{C}}   

\title{Quantum Time-Frequency Transforms}

\author{Mark Ettinger\,%
\thanks{\,\mbox{NIS--8}, \mbox{MS~B230}, 
Los Alamos National Laboratory, Los Alamos, NM~87545, USA.
Email: \texttt{\boldmath ettinger$\mathchar"40$lanl.gov}.}\\%
{\protect\small\sl Los Alamos National Laboratory\/}}

\date{}

\begin{document}

\maketitle

\begin{abstract}
Time-frequency transforms represent a signal as a mixture of its time domain
representation and its frequency domain representation.  We present efficient
algorithms for the quantum Zak transform and quantum 
Weyl-Heisenberg transform.
\end{abstract}

\section{Introduction}
The Fourier transform is an operator that expresses a time-dependent
signal as a sum (or integral) of periodic signals.  In other words the
Fourier transform changes a function of time $s(t)$ into a function
of frequency $S(\omega)$.  If a signal is a function of time it said to
be in the ``time domain'' and if it is a function of frequency it is said
to be in the ``frequency domain''.  For signals whose spectrum is changing
in time, i.e. nonstationary signals, sometimes the best description is a
mixture of the time and frequency components.  Signal representations which
mix the time and frequency domains are called, naturally enough,
``time-frequency representations'' and are often used to describe 
time-varying signals for which the pure frequency or Fourier representation 
is inadequate \cite{C95},\cite{QC96}.  A familiar example of a 
time-frequency representation is a musical score, which describes {\em when}
(time) certain {\em notes} (frequency) are to be played.

Formally speaking for our present purposes, a quantum signal is 
simply a quantum state $\ket{\psi}$
where the Hilbert space is the group algebra $\complex[G]$ of a finite
abelian group $G$.  
The Quantum Fourier Transform (QFT) is central to the important quantum 
algorithms for factoring and discrete logarithm.  Mathematically speaking,
the Quantum Fourier Transform 
is a linear operator on the Hilbert Space $\complex[G]$ which is a change of basis
from the basis of group elements $\{\ket{g_1},....,\ket{g_{|G|}}\}$
to the basis of characters of $G$, $\{\ket{\chi_1},\ket{\chi_2},...,
\ket{\chi_{|G|}}\}$.

We present efficient algorithms for quantum versions
of the Zak and Weyl-Heisenberg
transforms.  Both these time-frequency tranforms 
can be seen as generalizations of Fourier transforms
and the quantum 
algorithms make heavy use of the Quantum Fourier Transform.  We
follow the theory and notation of \cite{TA98} and recommend this book
as background to this material.

\section{Zak Transforms}
\subsection{Background}
Let $A$ be a finite, abelian group, $A^*$ the group of characters of
$A$ (note: in this paper $^*$ does {\em not} mean conjugation), $B \leq A$ a subgroup of $A$, \linebreak $B_* = \{ a^* \in A^*: a^*(b) = 1, b \in B\}$
the dual to $B$, and $f \in C[A]$, the group algebra of $A$.  Define
$$Z(B)f \in C[A \times A^*]$$ by the formula
$$Z(B)f(a,a^*) = \sum_{b \in B}f(a + b)\overline{a^*(b)}.$$
$F = Z(B)f$ is called the Zak transform of $f$ over $B$.  A simple calculation
shows that $F(a + b,a^* + b_*) = a^*(b) F(a,a^*)$
where $b \in B$ and $b_* \in B_*$.  Therefore
$F$ is determined by its values on a set of coset representatives of
$B \times B_*$ in $A \times A^*$ and thus conceptually we may think
of $F$ as a function on $T$ where $T$ is a set of coset representatives.  
Since 
$$\frac{|A \times A^*|}{|B \times B_*|} = \frac{|A|^2}{|B||B_*|} = |A|$$ we have the same number of degrees of freedom with 
which we started.  Notice that if $B$ contains only the identity, i.e.
is the trivial subgroup, then $Z(B)f(a,a^*) = f(a)$ and is basically
the identity map.  Also notice that if $B = A$ then 
$Z(A)f(0,a^*) = \braket{a^*}{f}$ and
therefore $Z(A)f$ is basically the Fourier transform of $f$.  So the
Zak transform mediates between the time domain and frequency domain depending
on the subgroup $B$.

Consider the function $f(a_0) = \delta(x-a_0)$ which is $1$ on $a_0$ and
$0$ otherwise.  Applying the above formula for the Zak transform
yields \linebreak
$F(a,a^*) = a^*(a - a_0)$ for $a \in a_0 + B, a^* \in A^*$ and
$0$ otherwise.  But since $F$ is determined by its values on a set
of coset representatives of $B \times B_*$ in $A \times A^*$
let us introduce such a set of representatives $T = T_1 \times T_2
= \{(x_i,a^*_j)\}$
where $T_1 = \{x_i\}$ is a set of coset representatives of $B$ in $A$ and
$T_2 = \{a^*_j\}$ is a set of coset representatives of $B_*$ in $A^*$.
Bearing in mind the above transformation of a delta function, 
we now offer our definition of
 the Quantum Zak Transform (QZT) (with respect to $T$) by
$$\ket{a} \mapsto \frac{1}{\sqrt{|B|}}\sum_{a^*_j \in T_2} a^*_j(x_a - a)\ket{x_a}\ket{a^*_j}.$$
where $x_a \in T_1$ is the coset representative of $a$.  Now notice that
$x_a - a \in B.$  Therefore $a^*_j$ is restricted to $B$ and therefore
can be considered to be a character of $B$, i.e. an element of $B^*$, and
this restriction is independent of the choice of coset representative, i.e.
it is {\em natural} or {\em canonical}.
Therefore an equivalent formulation of the QZT is given by
$$\ket{a} \mapsto \frac{1}{\sqrt{|B|}}\sum_{b^* \in B^*} b^*(x_a - a)\ket{x_a}\ket{b^*}.$$  The only difference in these two formulations is in the
{\em interpretation} of the observed content of the second register.

\subsection{The Quantum Algorithm}
We now show that the QZT is efficiently implementable.  Define $P(B)$
to be the transform $$P(B)\ket{a} = \ket{x_a}\ket{x_a - a}$$ which
decomposes $a$ into its coset representative and the corresponding
element of $B$.  $P$ is clearly 
unitary and efficiently implementable.  After applying $P(B)$ we apply
the Quantum Fourier Transform (over the group $B$, denoted $F_B$) to the second
register.  This results in the state
$$\frac{1}{\sqrt{|B|}}\sum_{b^* \in B^*} b^*(x_a - a)\ket{x_a}\ket{b^*}.$$
Therefore the QZT is simply $Z(B) = (I \otimes F_B) \circ P(B).$

\section{Weyl-Heisenberg Transforms}
\subsection{Background}
Define $g_{(x,x^*)}(a) = g(a-x) x^*(a)$ to be the
{\em time-frequency} translate
of $g$ by $(x,x^*)$ where $g \in C[A]$.  We will use time-frequency 
translates to form orthonormal bases so we also require
$|g| = 1$.
Let $\Delta = B \times B_*$ and \linebreak $(g,\Delta) =
\{g_{(x,x^*)}: (b,b_*) \in \Delta\}.$  We call $(g,\Delta)$ a 
{\em W-H system over $\Delta$ with window $g$}.  A basic result
(\cite{TA98}, Theorem 12.1 {\em corrected} version) 
is that $(g,\Delta)$ is an orthonormal basis
of $C[A]$ if and only if for all $(a,a^*) \in A \times A^*$ we have
$|G(a,a^*)| = \sqrt{\frac{|B|}{|A|}}$ 
where the Zak tranform is taken over $B$.  
Because von Neumann measurements must be unitary we
will restrict our attention to window
functions $g$ which satisfy this constraint.  Utilizing POVMs one
could consider implementing nonorthonormal W-H systems but we will not
address this in this note.  This orthogonality constraint 
together with the earlier observation that $G$ is
determined by its values on a set of coset representatives of
$B \times B_*$ in $A \times A^*$
implies that orthonormal W-H systems are in bijective
correspondence with the set of all $|A|$-tuples of complex numbers with
modulus $\sqrt{\frac{|B|}{|A|}}$.  In this note we will restrict the W-H systems under
consideration by assuming that for each $(a,a^*) \in A \times A^*$ 
the phase of $G(a,a^*)$ is a rational fraction of $2 \pi$ 
which we can compute in polynomial
time.  Whether or not this last assumption is excessively 
restrictive would depend on the intended application.  Notice that
if $g$ is the constant function $g = \frac{1}{\sqrt{|A|}}$ and $\Delta = \{0\} \times A^*$ then $(g,\Delta)$ is the (normalized) 
Fourier basis, $G(a,a^*) = \frac{1}{\sqrt{|A|}}$ and this restriction holds
trivially.

We define the Quantum Weyl-Heisenberg Transform (QWHT) by
$$\ket{\psi} \mapsto \sum_{(b,b_*) \in \Delta}
  \braket{\psi}{g_{(b,b_*)}} \ket{b,b_*}.$$
In other words, the QWHT expresses $\ket{\psi}$ in the orthonormal basis
of time-frequency translates of the window function.

\subsection{The Quantum Algorithm}
Let $$f = \sum_{(b,b_*) \in \Delta}\alpha(b,b_*)g_{(b,b_*)}$$ i.e. the 
$\alpha$'s are the coeffients of the WH-expansion of $f$. Define
$$P(a,a^*) = \sum_{(b,b_*) \in \Delta} \alpha (b,b_*) b_*(a) 
\overline{a^*(b)}.$$
Notice that $P$ is $\Delta$-periodic and that the $\alpha$'s are, by
definition, the Fourier coefficents (over $A \times A^*$) of $P$.  
A fundamental result
(\cite{TA98}, Theorem 7.5) states that $F = GP$.  This result suggests
an algorithm for computing the WH-coefficients of $f$, namely compute
the Fourier coefficients of $P = \frac{F}{G}.$

Define $\Phi(g)$ to be the unitary transformation which acts on the
Hilbert space $C[T]$ (recall $T$ is the set of coset representatives
of $B \times B_*$ in $A \times A^*$) by
$$\ket{x_i}\ket{a^*_j} \mapsto \frac{1}{G(x_i,a^*_j)} \ket{x_i}\ket{a^*_j}.$$  Since the phase of $G(x_i,a^*_j)$ is, by assumption, a rational fraction
of $2 \pi$ computable in polynomial time
we may efficiently implement $\Phi(g)$ by the phase kickback technique
described in \cite{Cleve}.  Finally in order to complete our description
of the algorithm, we must assume that we are given an explicit isomorphism
between $A$ and $A^*$.  These groups are isomorphic, though not canonically
so.  Therefore in any computational situation we provide an explicit 
isomorphism by choosing an explicit computational representation of the groups
$A$ and $A^*$.  This isomorphism induces explicit isomorphisms between
$B$ and $B^*$ and between the factor group $A/B$ and $B_*$.  We will see
shortly how we will employ these three interrelated isomorphisms.  
We will highlight this interrelation, and abuse notation,
by using the symbol $\phi$ to refer to all three of these isomorphisms,
allowing for context to make the usage clear.  As in the case of the
Zak transformtion, these isomorphisms are simply {\em reinterpretations} of
the contents of the registers.

Our QWHT is the sequence $F_{B_* \times B} \circ \Phi(g) \circ Z(B).$  Let us
see how this unitary transformation acts on $\ket{a}$.  We have
$$Z(B)\ket{a} = \frac{1}{\sqrt{|B|}}\sum_{a^*_j \in T_2} a^*_j(x_a - a)\ket{x_a}\ket{a^*_j}$$ and
then after applying $\Phi(g)$ we obtain:
$$\frac{1}{\sqrt{|B|}}\sum_{a^*_j \in T_2} \frac{a^*_j(x_a - a)}{G(x_a,a^*_j)}\ket{x_a}\ket{a^*_j}$$
which by the fundamental result discussed above equals:
$$\frac{1}{\sqrt{|B|}}\sum_{b^*} P(x_a,b^*) \ket{x_a}\ket{b^*}$$  where we are now considering
the contents of the second register to be an element of $B^*$.  We now 
utilize our explicit isomorphisms to reinterpret the contents of the first
register as an element of $B_*$ and the contents of the second register
as an element of $B$:
$$\frac{1}{\sqrt{|B|}}\sum_{b} P(b_*,b) \ket{b_*}\ket{b} = 
\frac{1}{\sqrt{|B|}}\sum_{\phi(a^*_j)} P(\phi(x_a),\phi(b^*)) \ket{\phi(x_a)}\ket{\phi(b^*)}.$$
By applying the final transformation in the sequence
$F_{B_* \times B}$ we obtain our desired expansion:
$$\sum_{(b,b_*) \in \Delta} \braket{a}{g_{(b,b_*)}}\ket{b}\ket{b_*}.$$

\section*{Acknowledgements}
We thank Myoung An, Richard Cleve, Peter Hoyer, Michele Mosca, and
Richard Tolimieri for helpful conversations.

\end{document}